
\documentclass{article}
\usepackage{setspace}
\usepackage{arxiv}
\usepackage{caption}
\usepackage{subcaption}
\usepackage[dvipsnames]{xcolor}
\usepackage{amsmath}
\usepackage[utf8]{inputenc} 
\usepackage[T1]{fontenc}    
\usepackage{hyperref}       
\usepackage{url}            
\usepackage{booktabs}       
\usepackage{amsfonts}       
\usepackage{nicefrac}       
\usepackage{microtype}      
\usepackage{lipsum}
\usepackage{graphicx}
\usepackage[numbers]{natbib}
\usepackage{bbm}
\graphicspath{ {./images/} }

\usepackage{enumitem}
\setlist{nosep}

\newcommand{\R}{\mathbb{R}}
\renewcommand{\l}{\big(}
\renewcommand{\r}{\big)}
\newcommand{\minimize}{\operatornamewithlimits{minimize}}

\newcommand{\T}{\mathsf{T}}
\newcommand{\indicator}[1]{\mathbf{1}\left\{#1\right\}}

\title{Smooth multi-period forecasting with application to prediction of COVID-19 cases}

\author{
 Elena Tuzhilina \\
  Department of Statistics\\
  Stanford University\\
  \texttt{elenatuz@stanford.edu} \\
  \And
  Trevor J. Hastie \\
  Department of Statistics and \\
  Department of Biomedical Data Science\\
  Stanford University\\
  \texttt{hastie@stanford.edu} \\
     \And
  Daniel J. McDonald\\
  Department of Statistics \\
  The University of British Columbia \\
  \texttt{daniel@stat.ubc.ca}
    \And
     J. Kenneth Tay \\
  Data Science and Applied Research\\
  LinkedIn Corporation \thanks{J. Kenneth Tay was a part of Department of Statistics at Stanford when the project was completed.}\\
  \texttt{kjytay@alumni.stanford.edu}
  \And
  Robert  Tibshirani \\
  Department of Statistics and \\
  Department of Biomedical Data Science\\
  Stanford University\\
  \texttt{tibs@stanford.edu}
}

\begin{document}

\maketitle
\begin{abstract}
Forecasting methodologies have always attracted a lot of attention and have become an especially hot topic since the beginning of the COVID-19 pandemic. 
In this paper we consider the problem of multi-period forecasting that aims to predict several horizons at once. We propose a novel approach that forces the prediction to be "smooth" across horizons and apply it to two tasks: point estimation via regression and interval prediction via quantile regression. This methodology was developed for real-time distributed COVID-19 forecasting. We 
illustrate the proposed technique with the CovidCast dataset as well as a small simulation example.
\end{abstract}

\allowdisplaybreaks

\section{Introduction}
\label{intro}
Time series forecasting techniques are used to predict events that occur over time by analyzing trends and patterns in past data. They are widely applicable across many fields of including finance, economics, politics, sports, meteorology and epidemiology. The latter area became especially important since the beginning of the COVID-19 pandemic in December 2019.  

Several time-series forecasting techniques have been proposed in the literature. Standard statistical methods based on regressive models such as \textit{autoregressive (AR)}, \textit{moving average (MA)}, \textit{autoregressive moving average (ARMA)}, \textit{autoregressive integrated moving average (ARIMA)} have been commonly used to forecast time-series (see, for example, \citep{boxjen76}). These Box-Jenkins
methods are  particularly efficient when applied to a linear stationary time series; they can accommodate the non-linear case by applying some appropriate transformation first. 
More recent approaches are based on machine learning methods, in particular, artificial neural networks (see, for example, \citep{zhang1998,nesreen2010,makridakis2018}). Compared to the ARIMA-type models, these often demonstrate better performance in forecasting non-linear signals.

The standard application of these techniques aims to predict the signal for a single forecast horizon (or "ahead"), most often one-step-ahead. However, in some applications such as epidemiology, where decisions are often based on the future trend of signal, simultaneous forecasts for multiple aheads can be of great interest. 
One of the popular methods for predicting several ahead values is \textit{multi-stage prediction (MSP)} (see, for example \citep{chen2004}) or \textit{multi-period forecasting (MFP)}. This approach is usually based on a single output model which is applied recursively, i.e. the predicted value of the signal three weeks ahead is determined based on the already-produced predicted values for one and two weeks ahead. The main disadvantages of such an iterative procedure is error propagation. An alternative method suggested in the literature is called the \textit{multiple-input multiple-output approach (MIMO)}, which aims to predict
a vector of future values all at once (see, for example, \citep{bontempi2008,taieb2010}). Detailed comparisons between different MIMO techniques can be found in \citep{cheng2006} and \citep{an2015}.

In this study we introduce a novel approach for predicting multiple ahead values simultaneously which is based on the idea that the future signal can be well-approximated by a smooth curve. The rest of the paper is organized as follows. In Section \ref{forecasting} we introduce the general multi-period forecasting problem. 
In Sections \ref{baseline}--\ref{smoothing} we describe two regression-based approaches for solving it:
\begin{itemize}
    \item a simple baseline method that predicts all aheads independently of each other (often termed ``direct'' forecasting);
    \item and a novel MPF method that enforces smoothness across aheads.
\end{itemize}
We extend the methodology to the case that some of the response signals are unobserved in Section~\ref{missing} and propose an analogue based on quantile-regression in Section \ref{qr}. 
Sections \ref{simulation}, \ref{realdata} and \ref{qr:realdata} illustrate the MPF technique on a small simulation example as well as real COVID-19 case incidence data obtained from the Delphi Epidata CovidCAST API~\citep{Reinhart2021}. We conclude the paper with a Discussion where we suggest some future research directions. 

\section{Forecasting problem}
\label{forecasting}

In this section we state the general multi-period forecasting problem. The  aim is to predict multiple future values of a time-dependant variable using a set of features (also depending on time).
We begin by introducing some notation. 
Suppose that we measure a response variable $Y_{i}(t)$ and a vector of $p$ covariates $X_{i}(t) = \l X_{i1}(t), \ldots, X_{ip}(t)\r $ at time $t$ and location $i$. 
Denote by $A = \{a_1,\ldots,a_q\}\in\R^q_{\geq 0}$ the sorted set of target ahead values for the response variable; $L_k = \{\ell_{k1},\ldots,\ell_{km_k}\}\in\R_{>0}^{m_k}$ a set of ``lags'' for the $k$-th predictor; and $L = \{L_1, \ldots, L_p\}$ a list of lags for all the covariates. 
Then the goal of \textit{multi-period forecasting (MPF)} is to predict the response variable for all the aheads, i.e. 
$$Y_{i}(t+A) = \l Y_{i}(t+a_1), \ldots, Y_{i} (t+a_q)\r\in\R^q ,$$ using all the lagged features at location $i$, i.e.
$$X_{i}(t-L) = \l X_{i1} (t-L_1), \ldots, X_{ip}(t-L_p)\r\in\R^{m}.$$
Here, by analogy with the response,
$$X_{ik}(t-L_k) = \l X_{ik}(t-\ell_{k1}), \ldots, X_{ik}(t-\ell_{km_k})\r\in\R^{m_{k}}$$ 
represents the lagged values of the $k$-th predictor at location $i$ and $m=\sum_{k=1}^{p}m_k$ corresponds to the total number of lagged predictors.

A simple example of an MPF problem is: on December 15, predict the expected number of newly reported of COVID-19 cases on December 15 and December 22 using the number of visits to the doctor on December 8 and December 1 across all the U.S.\ states. In this case,
\begin{itemize}
    \item $t$ is December 15, the forecast date; 
    \item $i$ represents a U.S.\ state;
    \item $Y_{i}(t)$ is the number of COVID-19 cases in state $i$ on day $t$;
    \item $X_i(t) = \l X_{i1}(t)\r$ represents the number of doctor visits in state $i$ on day $t$;
    \item $A = \{0, 7\}$ is the set of ahead values; 
    \item $L_1 = \{7, 14\}$ is the set of lags. 
\end{itemize}

Note that in many applications the response variable is also included in the set of predictors, thereby incorporating the historical values of the response into the feature set.

\section{Baseline linear model}
\label{baseline}
A straightforward (direct) multi-period forecaster is a linear model for each location $i$, timestamp $t$ and ahead value~$a$: 
\begin{align}
Y_{i}(t + a) = \sum_{k=1}^p\sum_{\ell \in L_k}X_{ik}(t-\ell)~b_{k \ell}(a) + \epsilon_i(t+a).
\label{eq:mrr:model}
\end{align}
Here $\epsilon_{i}(t+a)\sim\mathcal{N}(0, \sigma^2)$ are i.i.d errors and $b_{k \ell}(a)$ are unknown model coefficients.
In what follows, we assume that the measurements are done at $n$ locations and that multiple past values are available. If we denote the set of the available past timestamps by $T = \{t_1, \ldots, t_N\}$
then model (\ref{eq:mrr:model}) leads us to the following objective
\begin{align}
    \sum_{i = 1}^n \sum_{t\in T} \sum_{a \in A}\left(Y_{i}(t + a) - \sum_{k=1}^p\sum_{\ell \in L_k}X_{ik}(t-\ell)~b_{k \ell}(a)\right)^2
    \label{eq:mrr:loss}
\end{align}
that we aim to minimize w.r.t.\ the model coefficients. We note that the resulting optimization goal is nothing but a multivariate least-squares problem: the loss is separable in terms of ahead values, so $b_{k\ell}(a)$ can be found independently for each $a\in A$ via ordinary least squares with response $Y_i(t+a)$ and predictors $X_i(t-L)$. 

For convenience we will restate the objective in matrix form.
To do so, we first denote all the coefficients corresponding to the $k$-th predictor by
$$b_{k}(a) = \l b_{k\ell_{k1}}(a), \ldots, b_{k\ell_{km_k}}(a)\r\in\R^{m_k}$$ 
and form the coefficient matrix  
$$
B = \begin{pmatrix}
b_1(a_1) & \cdots & b_p(a_1)\\
\vdots & \ddots & \vdots\\
b_1(a_q) & \cdots & b_p(a_q)\\
\end{pmatrix}\in\R^{q\times m}.
$$
Next, we denote the matrices of the response and the predictors measured at time $t$ by 
$$Y(t) = \begin{pmatrix}Y_{1}(t+A)\\
\vdots\\
Y_{n}(t+A)
\end{pmatrix}\in\R^{n\times q} \quad\mbox{and}\quad
X(t) = \begin{pmatrix}X_{1}(t-L)\\
\vdots\\
X_{n}(t-L)
\end{pmatrix}\in\R^{n\times m}$$ 
and concatenate all the data rowwise into
$$Y = \begin{pmatrix}Y(t_1)\\
\vdots\\
Y(t_N)
\end{pmatrix}\in\R^{Nn\times q} \quad\mbox{and}\quad
X = \begin{pmatrix}X(t_1)\\
\vdots\\
X(t_N)
\end{pmatrix}\in\R^{Nn\times m}.$$ 

Hence, the  MPF optimization problem in Equation~\ref{eq:mrr:loss} can be stated in multi-response regression (MRR) form~as 
\begin{align}
    \minimize_{B\in\R^{m\times q}}\|Y-XB^\T\|^2_F,
\label{op:mrr:mat}
\end{align}
where $\|Z\|^2_F = \sum _{ij} Z_{ij}^2$ is the squared Frobenius norm of a matrix $Z$. The explicit solution can be found via the formula 
$$\widehat B^\T = (X^\T X)^{-1}X^\T Y.$$ 
We will refer to this forecaster as the Baseline MPF.

\section{Smoothing constraint}
\label{smoothing} 

The main disadvantage of the Baseline model (\ref{op:mrr:mat}) is that the coefficients for all the response columns are computed independently of each other. 
In other words, the model completely ignores the underlying data structure, i.e. that each column of $Y$ represents the same signal measured for different ahead values.
To incorporate this information into the MPF problem we desire some smoothness in the model coefficients. 

Specifically, we desire that each $b_{k\ell}(a)$ is a smooth function of ahead values. Such smoothness can be enforced by requiring $B$ to be representable as a linear combination of smooth basis functions $h_1(a),\ldots,h_d(a)$ (e.g. a spline or polynomial). This suggests the representation
\begin{align}
b_{k\ell}(a) = \sum_{j=1}^d\theta_{jk\ell}h_j(a) \mbox{ for some } \theta_{jk\ell}\in\R.
\label{eq:smooth}
\end{align}
Here $d$ is a hyperparameter that controls the flexibility of $b_{k\ell}(a)$. In what follows, we refer to $d$ as the \textit{degrees-of-freedom}. Combining (\ref{eq:mrr:loss}) with (\ref{eq:smooth}) leads us to the smooth multi-period forecasting (SMPF) objective
\begin{align}
   \minimize_{\theta_{jk\ell},\ \forall j,k,\ell} \sum_{i = 1}^n \sum_{t\in T} \sum_{a \in A}\left(Y_{i}(t + a) - \sum_{k=1}^p\sum_{\ell \in L_k}X_{ik}(t-\ell)\sum_{j=1}^d\theta_{jk\ell}h_j(a)\right)^2.
\label{eq:mpf:loss}    
\end{align}
Note that the second term in (\ref{eq:mpf:loss}) 
involves all the unknown parameters $\theta_{jk\ell}$ of the model, so the resulting loss function is no longer separable. However, since the predicted values
\begin{align}
\hat Y_{i}(t + a) = \sum_{k=1}^p\sum_{\ell \in L_k}X_{ik}(t-\ell)\sum_{j=1}^d\theta_{jk\ell}h_j(a)    
\label{eq:fit}
\end{align}
is a linear function of the coefficients it is still possible to find the explicit solution via regression.

Again, it is convenient to rewrite the loss function in matrix form. To do so, we first store all the coefficients in a matrix
\begin{align*}
    \Theta = \begin{pmatrix}\theta_{11}&\ldots&\theta_{1p}\\\ldots&\ldots&\ldots\\\theta_{d1}&\ldots&\theta_{dp}\end{pmatrix}   
\in\R^{d\times m}, \mbox{ where } \theta_{jk}=(\theta_{jk\ell_{k1}}, \ldots, \theta_{jk\ell_{km_k}})\in\R^{m_k}.
\end{align*}
Next, we introduce the basis matrix 
\begin{align*}
 H = \begin{pmatrix}
 h_1(a_1) & \ldots & h_d(a_1)\\
 \ldots & \ldots & \ldots\\
 h_1(a_q) & \ldots & h_d(a_q)
\end{pmatrix}\in\R^{q\times d},
\end{align*}
where each column represents a function from the basis evaluated at all ahead values in $A$. As a result, one can restate constraint (\ref{eq:smooth}) in matrix form as $B = H\Theta$ and, together with (\ref{op:mrr:mat}), this implies the SMPF optimization can be written as
\begin{align}
    \minimize_{\Theta\in\R^{d\times m}}\|Y-X\Theta^\T H^\T\|^2_F.
\label{op:mpf:mat}
\end{align}
Note that in this problem the basis $H$ is considered to be fixed, so the only unknown parameter is $\Theta$. The degrees-of-freedom $d$, which controls the size of the basis, is the model's hyperparameter and can be chosen from a grid of values via cross-validation.

Similar to the baseline model, it is possible to find an explicit solution to (\ref{op:mpf:mat}). First, without loss of generality, we assume that $H$ has orthogonal columns. Otherwise, one can take the QR decomposition $H = QR$ and apply the change of variables $\widetilde H=Q$ and $\widetilde\Theta=R\Theta$. Next, since the Frobenius norm is invariant under orthogonal transformations we can restate problem (\ref{op:mpf:mat}) as 
\begin{align*}
    \minimize_{\Theta\in\R^{d\times m}}\|YH-X\Theta^\T \|^2_F,
\end{align*}
which is, again, a multi-response regression problem with solution
$$\widehat \Theta^\T  = (X^\T X)^{-1}X^\T YH.$$

\section{Missing values}
\label{missing}
This section extends the SMPF methodology proposed in Section \ref{smoothing} to the case when only part of the response matrix $Y$ is observed.
In forecasting applications, missing values often occur. For example, for a recent time $t$ and location $i$ we may not have observed response values $Y_i(t+a)$ for all ahead values $a\in A$ as some of them have not occurred yet. 
Moreover, the data can be updated at different times for different locations; thus, $Y_i(t+a)$ may not have been collected yet for some $i$.

To handle unobserved values we allow the set of ahead values to depend on the timestamp $t$ and location $i$ and denote it by $A_i(t)$. We also assume that each $A_i(t)$ is a subset of original $A=\{a_1,\ldots,a_q\}$. One can derive the new loss function as follows
\begin{align}
    &\sum_{i = 1}^n \sum_{t\in T} \sum_{a \in A_i(t)}\left(Y_{i}(t + a) - \sum_{k=1}^p\sum_{\ell \in L_k}X_{ik}(t-\ell)\sum_{j=1}^d\theta_{jk\ell}h_j(a)\right)^2.
\label{eq:mpf:loss:weight}
\end{align}

Similar to Sections \ref{baseline}--\ref{smoothing}, it is not hard to restate the SMPF optimization problem in matrix form. Defining
$$W_i(t + a)=\begin{cases}1 &\mbox{ if } a\in A_i(t),\\0 &\mbox{ otherwise,}\end{cases}$$ 
to be a binary weight matrix representing the missingness of the response, then
minimizing Equation~\eqref{eq:mpf:loss:weight} is equivalent to solving
\begin{align}
    \minimize_{\Theta\in\R^{d\times m}}\|W \circ (Y-X\Theta^\T H^\T )\|^2_F,
\label{op:mpf:mat:weight}
\end{align}
where $\circ$ refers to the element-wise Hadamard matrix product and $W$ is the matrix containing all the weights.

Unlike the unweighted case, weighted SMPF cannot be
reduced to a multi-response regression by simple manipulations with Frobenius norm.
However, since the second term in (\ref{op:mpf:mat:weight}) is a linear function of  $\Theta$ it is still possible to restate it as an expanded ordinary least squares problem. Denote $w, y\in\R^{Nnq}$ and $\theta\in\R^{dm}$ the vectors obtained by the concatenation of columns of matrices $W, Y$ and $\Theta^\T $, respectively. 
Writing $\widetilde X = H\otimes X$ as the Kronecker product between $H$ and $X$, then Equation~\eqref{op:mpf:mat:weight} is equivalent to solving
\begin{align}
    \minimize_{\theta\in\R^{dm}}\|w \circ (y-\widetilde X\theta)\|^2_2.
\label{op:smooth:weight:vec}
\end{align} 
Note that for general $w$ the solution can be found by means of the weighted regression with weights $w,$ response $y$ and feature matrix $\widetilde X.$ However, if the weights are binary one can simply remove the rows in $y$ and $\widetilde X,$ that correspond to the zero weights, and use simple linear regression.

\section{Simulation experiment}
\label{simulation}
In this section we test the SMPF model from Section \ref{missing} on a small simulation example. For simplicity we use only one forecast date $t$ and denote it as $t = 0.$ 
We fix the number of locations at $n = 1000$ and the number of predictors at $p=10.$ We also assume no lags for this model, i.e. $L_k =\{0\}$ for $k=1,\ldots,10.$ We first generate the matrix of covariates $X\in\R^{n\times p}$ with elements $X_{ik}\sim\mathcal{N}(0,1)$.
Further, we set the number of ahead values to $q = 30$ and the set of ahead values to $A = \{0,1,\ldots,29\}.$ To create $B$ we evaluate orthogonal quadratic polynomial basis at all elements in $A$ and store them column-wise as $H\in\R^{q\times d}$. Here $d = 3$ and each column of $H$ represents a basis function, including the intercept. Next, we draw the elements of the coefficient matrix $\Theta\in\R^{d\times m}$ from standard normal distribution. Finally, we generate the matrix of errors $E\in\R^{n\times q}$ with elements $\epsilon_{i,j}\sim\mathcal{N}(0, \sigma^2)$ and compute the response matrix as $Y = X\Theta^\T H^\T +E.$ We randomly sample $10\%$ of the $Y$ matrix elements and treat them as unobserved.

\begin{figure}[h!]
    \centering
    \includegraphics[width = \textwidth]{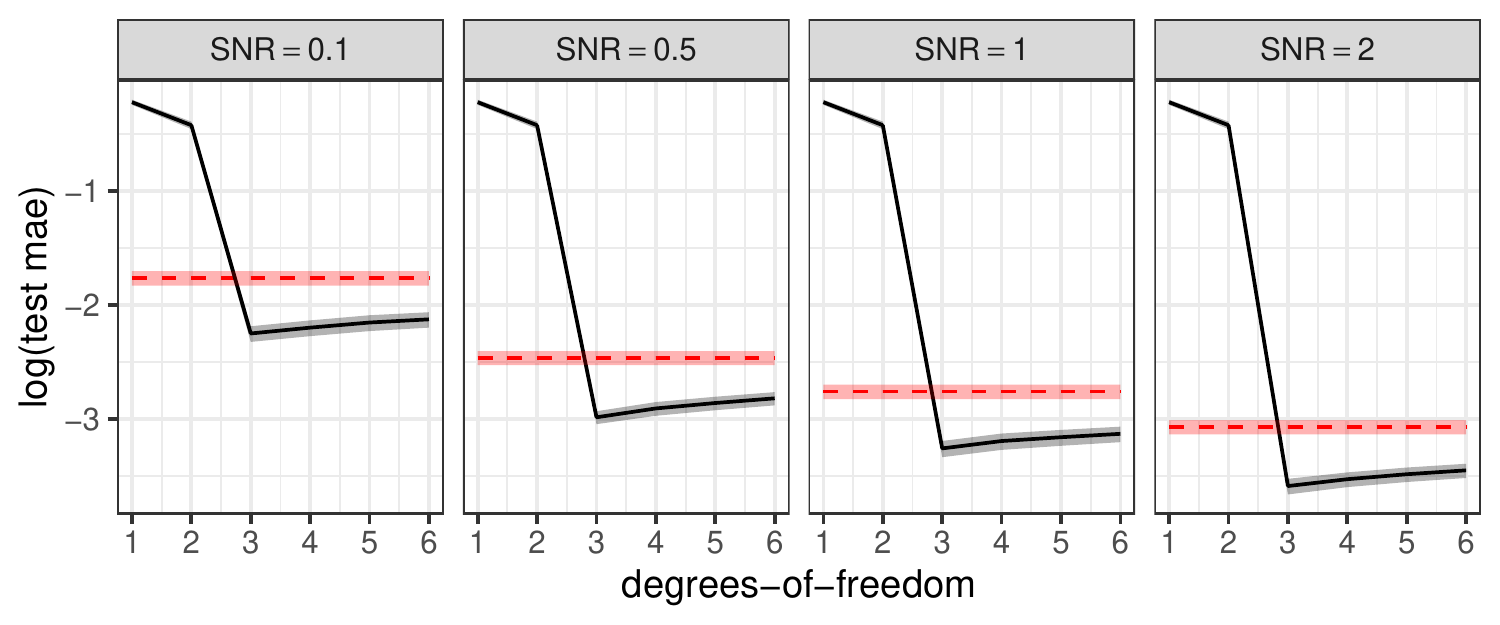}
    \caption{Simulation results. The solid black line represents the test MAE vs degrees-of-freedom computed by means of the smooth MPF model. The red dashed line corresponds to the MRR test score. Shaded regions represent 1SE interval computed across ten repeated simulations. Each panel corresponds to the simulated data with different SNR levels.}
    \label{fig:simulation}
\end{figure}

We use half of the locations to fit the smooth multi-period forecasting model and the remaining half to evaluate the model performance. We vary the error variance $\sigma^2$ such that the signal-to-noise ratio is ${\rm SNR}= 0.1, 0.5, 1, 2$, and we use mean absolute error (MAE) as the performance metric. Since, in practice, the true degrees-of-freedom is unknown, we it to vary over the grid $d = 1,2,\ldots,6$. For instance, $d=1$ corresponds to the ``null'' constant model and $d = 2$ represents straight line forecasts. Thus for each value of $\mathrm{SNR}$ we produce a curve (MAE vs. degrees-of-freedom).  The results are presented in Figure \ref{fig:simulation}, where we also add the baseline multi-response regression solution as a reference (dashed red line). 

According to the figure, for all ${\rm SNR}$ values the best smooth model outperforms the baseline, although the amount of improvement degrades slightly as ${\rm SNR}$ increases. Regardless of the signal-to-noise ratio, the minimum test score is achieved for the SMPF degrees-of-freedom 
around the true model value $d = 3$. Note that as the degrees-of-freedom increases, the SMPF still outperforms the Baseline, though setting $d=30$ would necessarily result in identical performance.
Therefore, in the simulation experiment the smooth multi-period forecasting model not only demonstrates the superior performance to the baseline method, but also is able to recover the true degrees-of-freedom.

\section{COVIDcast data experiments}
\label{realdata}

Now we apply the multi-period forecasting approaches on the real data obtained from the Delphi COVIDcast API~\citep{Reinhart2021}. This open-source data set, which is updated daily, tracks multiple signals related to the spread and impact of the COVID-19 pandemic across the United States on both county and state levels. It contains a wide variety of typical COVID-19 metrics such as incident cases, deaths, and hospitalizations, as well as many unique indicators derived from mobility data, internet symptom searches, healthcare utilization reports, and sample surveys.
For our experiments, we use three signals:
\begin{itemize}
\item \texttt{confirmed\_7dav\_incidence\_prop}: the daily number of new confirmed COVID-19 cases (computed per 100,000 people); 
\item \texttt{smoothed\_cli}: the estimated percentage of people with COVID-like illness, as measured by The Delphi Group at Carnegie Mellon University U.S. COVID-19 Trends and Impact Survey (CTIS), in partnership with Facebook~\citep{Salomon2021};
\item \texttt{smoothed\_hh\_cmnty\_cli}: the estimated percentage of people reporting illness in their local community, also measured by the Delphi US CTIS. 
\end{itemize}
The latter two indicators were obtained from a voluntary survey conducted by Facebook.
In order to reduce the weekly variability, all three signals are smoothed by taking the trailing average across a seven-day window. We consider the following forecast task:
\begin{itemize}
\item each location $i$ represents a U.S.\ county;
\item the response $Y_i(t)$ is the value of \texttt{confirmed\_7dav\_incidence\_prop} at county~$i$;
\item three predictive features are used, i.e. $X_i(t)=(X_{i1}(t), X_{i2}(t), X_{i3}(t))$ represents the values of \texttt{confirmed\_7dav\_incidence\_prop} as well as \texttt{smoothed\_cli} and \texttt{smoothed\_hh\_cmnty\_cli} at location $i$; 
\item ahead values $A = \{0,1,\ldots, 27\}$ target daily forecast targets over four weeks;
\item lag values $L = \{1,2,\ldots,28\}$ track the signal for four weeks preceding the forecast date.
\end{itemize}

\begin{figure}[tb!]
    \centering
    \includegraphics[width = \textwidth]{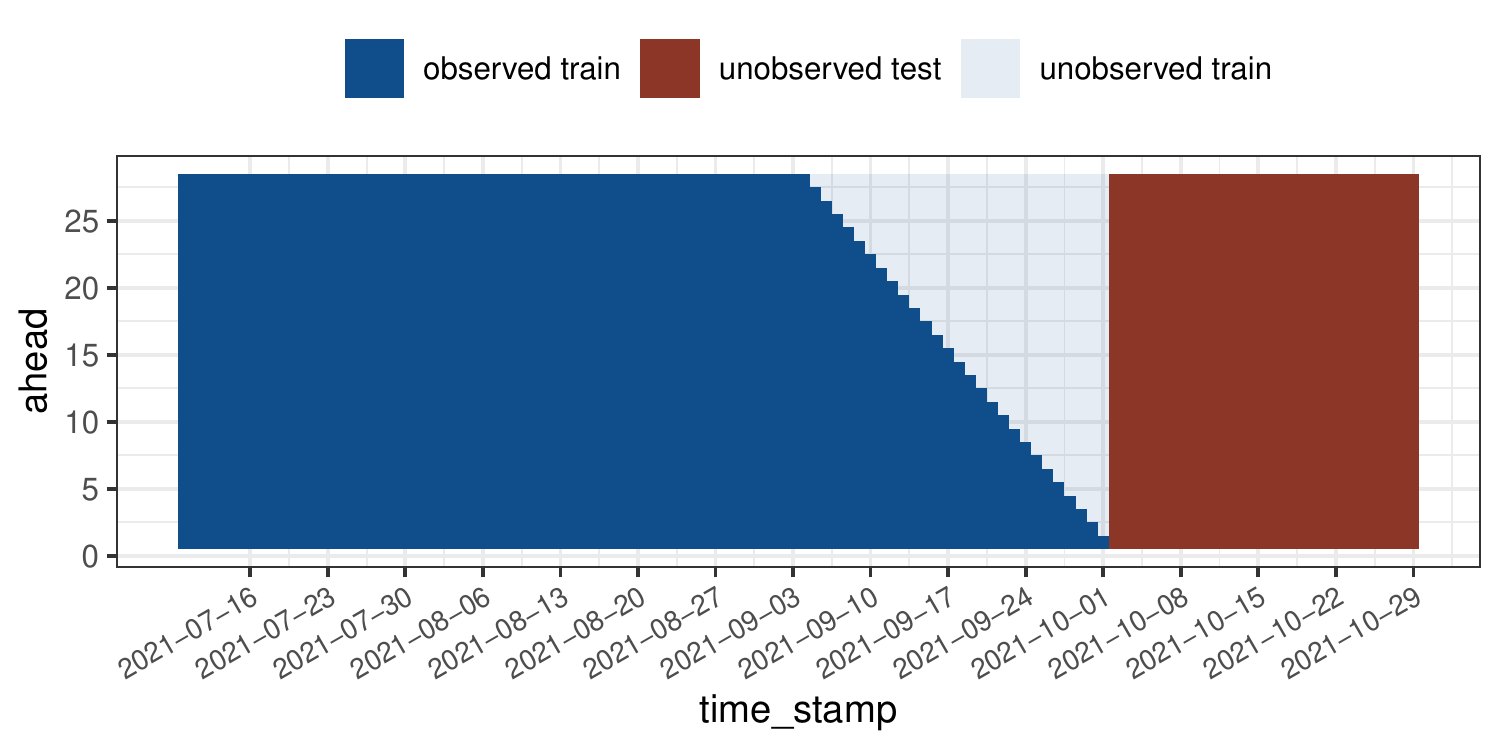}
    \caption{Schematic representation of missing values in response matrix when the ``as of'' date is set to October 1. Each column represents a timestamp; each row represents an ahead value; the element in row $a$ column $t$ corresponds to the $t+a$ time point. Blue and red colors represent train and test sets, respectively; light blue color corresponds the the time points after the ``as of'' date, which are treated as unobserved in the training phase. If $n=1$, i.e. only one location is considered, then the picture represents exactly the distribution of missing values in train $Y^\T $ (the blue block) joined with test $Y^\T $ (the red block).}
    \label{fig:missing:dist}
\end{figure}

The training set contains twelve weeks of daily data prior to 1 October 2021, that is
$$T_{train}=\{\mbox{10-Jul-2021, 11-Jul-2021, \ldots 1-Oct-2021}\}.$$ 
To make the experiment more realistic, the data was downloaded ``as reported on'' 1 October 2021, thereby making all the signals after this date to be unobserved. In other words, $Y_i(t+a)$ is unobserved, or equivalently, $W_i(t+a)=0$, if $t+a$ is any date after October 1. This practice also means that any revisions that would eventually be made after October 1 are not available.  The distribution of missing response values for the training set is shown in blue in Figure~\ref{fig:missing:dist}. To test both SMPF models with and without missingness (the solutions to Equations~\eqref{op:mpf:mat} and \eqref{op:mpf:mat:weight}) we explore two scenarios:
\begin{description}
\item[Scenario 1:] we remove all data for dates that would result in at least one unobserved ahead value, i.e. we use only the data from July 10 to September 4. In this case, the data is complete and we can use non-weighted SMPF for prediction. 
\item[Scenario 2:] we include all the data from July 10 to October 1. Since the response matrix is only partially observed, we fit the weighted modification of SMPF with binary weights. 
\end{description}

To make the solution more robust, among 581 counties with available survey data, we select the 300 with the highest average (across all the times) level of cases; we also remove all the observations containing missing values in the predictors. This results in $23079$ training observations and 84 predictors.

We fit both baseline and smooth MPF models on the training set.
For the smooth approach we use the orthogonal polynomial basis with intercept and vary the degrees-of-freedom in the grid $d=1, 2, \ldots, 6$. To evaluate the models' performance we download the response values for the same 300 counties and including four weeks of observations following October 1. In other words, the new dataset contains the timestamps
$$T_{test}=\{\mbox{2-Oct-2021, 3-Oct-2021, \ldots, 29-Oct-2021}\},$$
which results in $4780$ test observations. Since we are interested in estimating how well the model will do at forecasting the future cases, the test set is downloaded ``as of'' 27 January 2022
and therefore there are no missing responses. 


\begin{figure}[tb!]
    \centering
    \includegraphics[width = 0.9\textwidth]{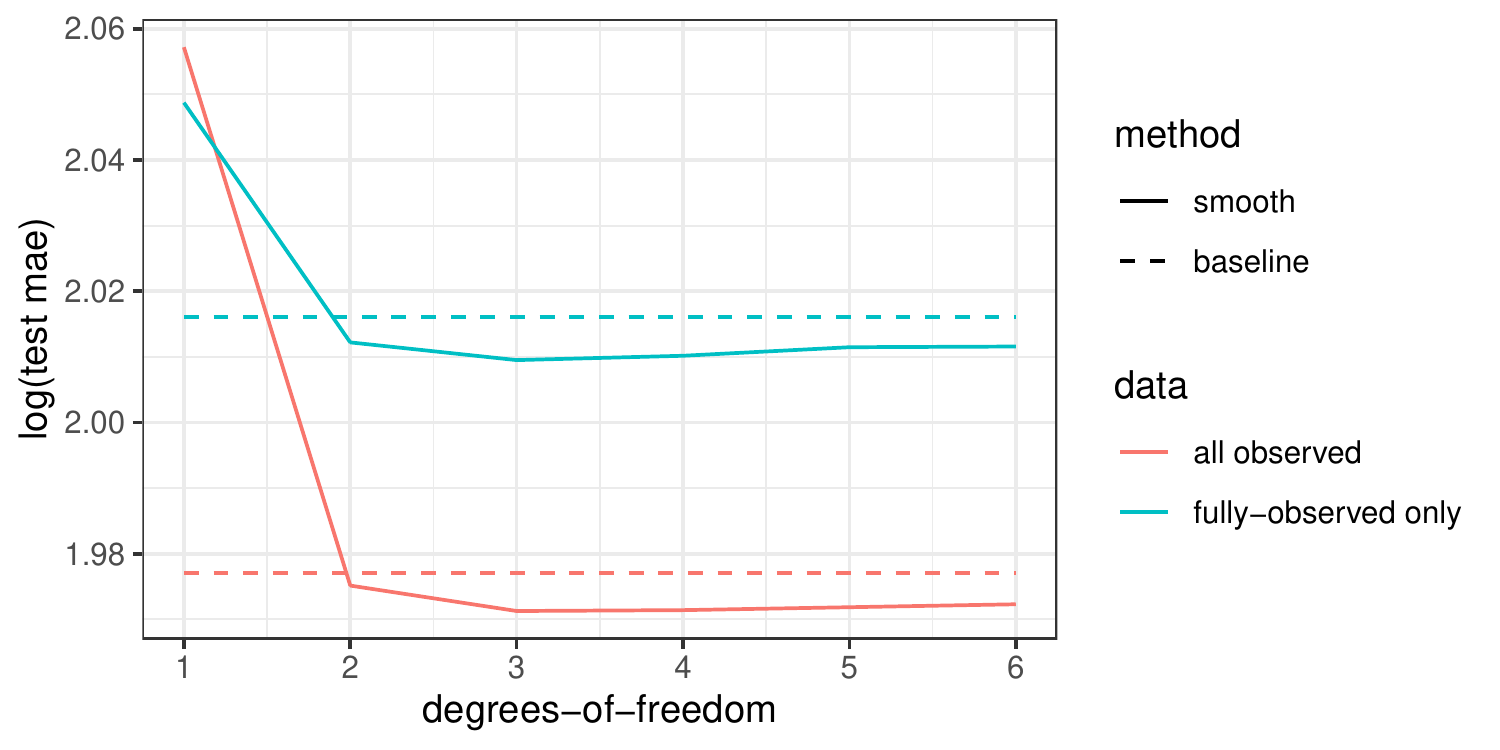}
    \caption{Comparing the test performance of the baseline and smooth MPF models while forecasting COVID-19. The data is downloaded ``as of'' October 1 and two scenarios are considered.
    Red color: the training data contains only the timestamps with fully-observed response vector (from July 10 to October 1), thus, the 
    response matrix
    has no missing values. Blue color: the response matrix includes all the available timestamps; thus, it has some missing values (blue curve). The solid line shows the test MAE scores computed for the smooth MPF models with different degrees-of-freedom, which vary in the grid $d = 1,2,\ldots, 6$. The dashed line represents the baseline model MAE. The plot demonstrates the superior performance of the smooth model to the baseline in both scenarios. 
    }
    \label{fig:mse:vs:df}
\end{figure}

\begin{figure}[p]
\centering

\includegraphics[width = 0.8\textwidth]{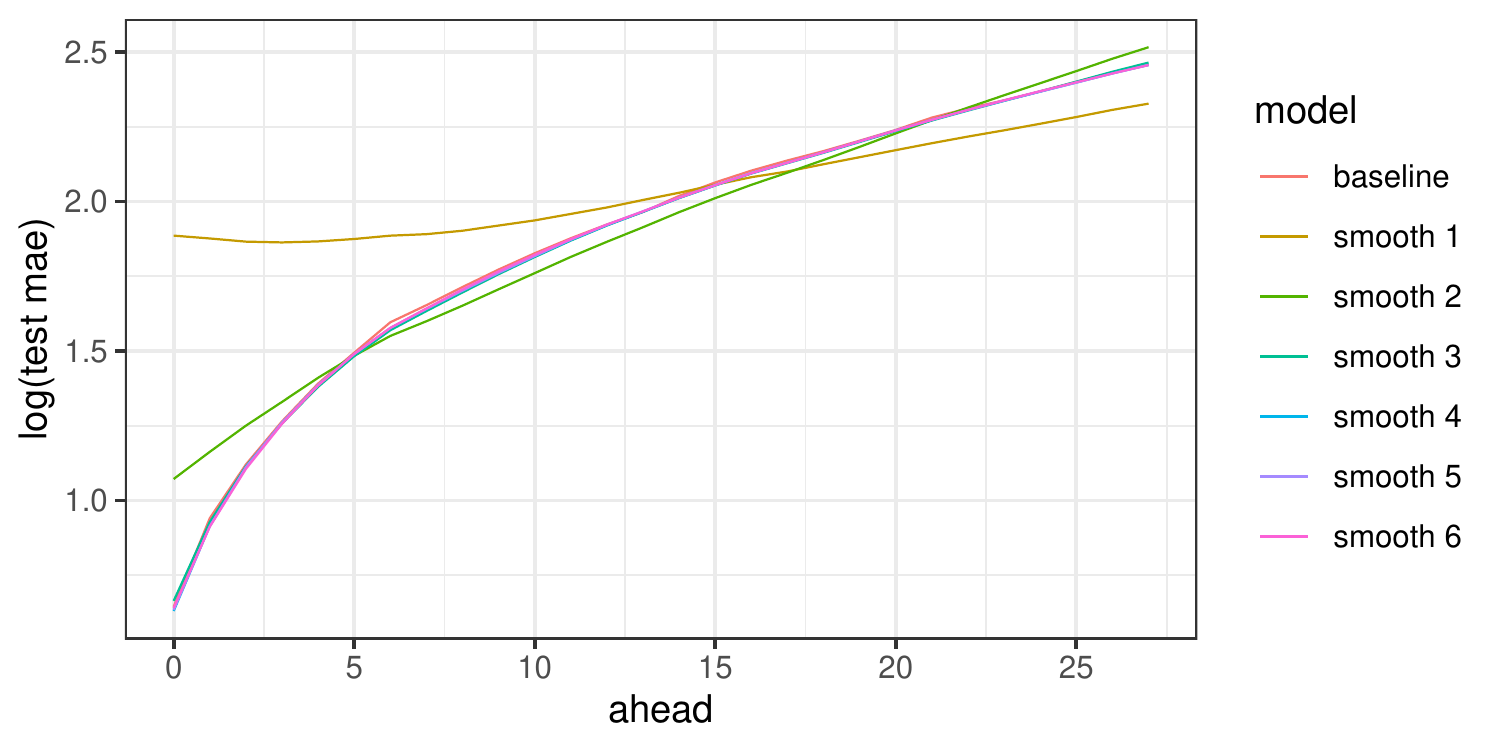}
\caption{Comparing the test performance of the baseline and smooth MPF models while forecasting COVID-19. The result is presented for the second scenario, i.e. when the all the timestamps from July 10 to October 1 are included even if the response vector is partially observed. In this plot the test MAE is calculated for each ahead value separately and each line corresponds to different models (either baseline or smooth with $d = 1,2,\ldots, 6$). The plot demonstrates that forecasting is more challenging for times which are further in the future. 
}
\label{fig:mse:vs:ahead}

\includegraphics[width=\linewidth]{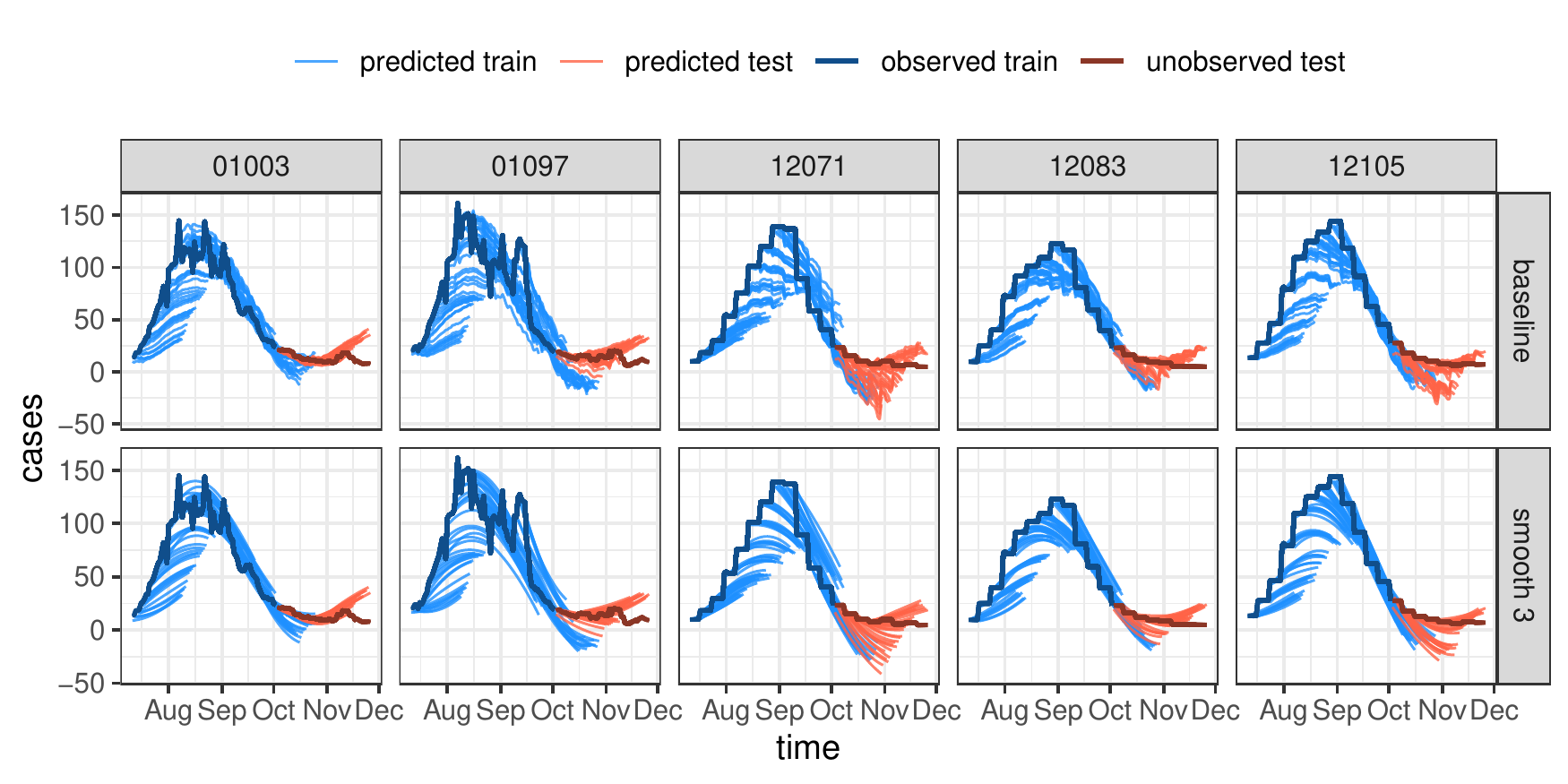}  
\caption{The plot displays the fits produced by the baseline and the optimal smooth model (with $d = 3$). The bold dark line shows the true value whereas predicted values are represented by bright thin lines (one line - one timestamp). Blue and red colors correspond to the train and test sets, respectively. The baseline MPF fit demonstrates irregular behavior which is moderated by smoothing.}
\label{fig:fits}
    
\end{figure}

In Figure \ref{fig:mse:vs:df} we show test mean absolute error (MAE) for smooth MPF models with different degrees-of-freedom (solid line). We also include baseline MAE as a reference (dashed line). Here, the test MAE is averaged across all the locations, timestamps and ahead values.
We start by comparing two data scenarios (blue and red colors in the figure).
According to the plot, using all the data available before the ``as of'' date implies better test performance. This can be explained by the fact that COVID data is quite volatile, so including more recent observations allows the model to more accurately predict the future trend. This, however, comes at a price of increased computational cost. For a fully-observed response matrix the solution can be found via pre-multiplying $Y$ by $H$ and fitting the multi-response regression with feature matrix $X\in\R^{Nn\times m}$ and response matrix $YH\in\R^{Nn\times d}$. At the same time, the partially observed case requires us to solve a much larger regression problem with feature matrix $H\otimes X\in\R^{Nnq\times md}$ and response $y\in\R^{Nnq}$. Next, by comparing the smooth and baseline MPF test scores we conclude that smoothing improves the performance of multi-period forecaster. From the red and blue curves in Figure \ref{fig:mse:vs:df} one can infer that, for both scenarios, the optimal value for the degrees-of-freedom is $d = 3$. The remaining results in this section are presented for the second data scenario, where the response matrix is partially observed.

To get more granular information on the model performance, we compute MAE separately for each column of $Y$ and plot the dependence of test error on the ahead value. In Figure \ref{fig:mse:vs:ahead} we observe that, as one would expect, the accuracy decreases for larger ahead values for all models under consideration. In other words, forecasting is more challenging for time points that are farther into the future. 

Finally, we compare baseline MPF with the best smooth model, i.e. the one that attains the lowest test score. Note that $d=3$ gives  quadratic dependence of the regression coefficients on time. Thus, the most promising approach is to predict some quadratic trend for cases at each timestamp. In Figure \ref{fig:fits} each thin bright line starts at a timestamp and represents the predicted cases for the coming four weeks (28 ahead values). Here, the top row shows the baseline predictions, and the bottom row corresponds to those obtained by the optimal smooth model. To visualize and compare the MPF performance on the train and test sets, we include both train (blue color) and test (red color) fits to the plot. We also add the ground truth cases as a reference (dark bold line). To make the figure more readable, we present the results only for the five counties with the highest average case values and display each county in a separate panel. By analyzing this plot, we can see that the baseline model produces fits which look more wiggly, or noisy, relative to the smooth MPF prediction. This extra noise in the regression coefficients results in higher test MAE of the baseline compared to the competitor. 
Note that when true cases are close to zero, MPF may predict (impossible) negative values. One can easily fix this either by taking a log-transform of cases or by imposing a constraint on the predicted values.

\section{Quantile forecasting}
\label{qr}

Now we shift from the point estimation task, which we handled by means of least squares regression, to interval prediction. In this section, we employ quantile regression (QR) to estimate intervals within which signals have a high probability of occurring (see, for example, \citep{koenker2005}).
We begin by introducing the the baseline quantile multi-period forecasting (QMPF) method.
For a quantile $\tau\in[0,1]$ consider the pinball loss function
$$\rho_\tau(y,\hat y) = \begin{cases}\tau(y-\hat y) & \mbox{ if } y \geq \hat y,\\
(1-\tau)(\hat y - y) & \mbox{ otherwise}.
\end{cases}$$
Then goal is to solve the following objective
\begin{align}
    \minimize_{b_{k\ell}(a)} \sum_{i = 1}^n \sum_{t\in T} \sum_{a \in A_i(t)}\rho_\tau\bigg( Y_{i}(t + a),~\sum_{k=1}^p\sum_{\ell \in L_k}X_{ik}(t-\ell)~b_{k \ell}(a)\bigg).
\label{eq:qmpf:loss}
\end{align}
Note that the above optimization task is stated in general form, where the set of ahead values can vary for each timestamp $t$ and location $i$. We again assume $A_i(t)\subseteq A.$ 

Similar to Section \ref{missing}, the solution to the QMPF problem can be found separately for each ahead value. Namely, for each $a$ it amounts to fitting quantile regression with feature matrix $X$ and the response vector which includes all the observed elements from $Y$ that corresponds to $a$. As a result, each ahead value can be handled very efficiently by linear programming methods (see, for example the software \citep{koenker2004}).

Incorporating the smoothness into the coefficients leads us immediately to the smooth version of the QMPF objective
\begin{align}
    \sum_{i = 1}^n \sum_{t\in T} \sum_{a \in A_i(t)}\rho_\tau \bigg( Y_{i}(t + a),~ \sum_{k=1}^p\sum_{\ell \in L_k}X_{ik}(t-\ell)\sum_{j=1}^d\theta_{jk\ell}h_j(a)\bigg),
\label{eq:sqmpf:loss}    
\end{align}
which we aim to minimize w.r.t.\ $\theta_{jk\ell}.$
By analogy with Section \ref{missing}, the smooth problem can be reduced to fitting a weighted QR through some simple manipulations with $X, Y, H$ and $\Theta$. Specifically, one can show that minimizing (\ref{eq:sqmpf:loss}) is equivalent to solving
\begin{align}
    \minimize_{\theta\in\R^{d m}}\sum_{i=1}^{Nnq} w_{i} \cdot \rho_\tau\l y_{i},\ \widetilde X_{i}^\T \theta\r.
\label{op:sqmpf:mat:weight}
\end{align} 
Here, $y, w\in\R^{Nnq}$ and $\theta\in\R^{dm}$ correspond to the vectors obtained by the concatenation of columns of matrices $Y, W$ and $\Theta^\T $, respectively; $W$ is the matrix of binary weights representing the the missing responses in $Y$; and 
$\widetilde X_i$ is the $i$-th row of $\widetilde X = H\otimes X$.

Note that, unlike the multiple least squares case, where the computations can be significantly simplified for fully-observed responses by pre-multiplying $Y$ by $H$, the QR loss is not invariant under the orthogonal transformations. Thus, computing the extended feature matrix $\widetilde{X}$ is necessary for the smooth QMPF technique, regardless of the missingness pattern.

\section{Quantile forecasting in COVIDcast study}
\label{qr:realdata}

\begin{figure}[p]
\begin{subfigure}{0.88\textwidth}
  \centering
  \includegraphics[width=\linewidth]{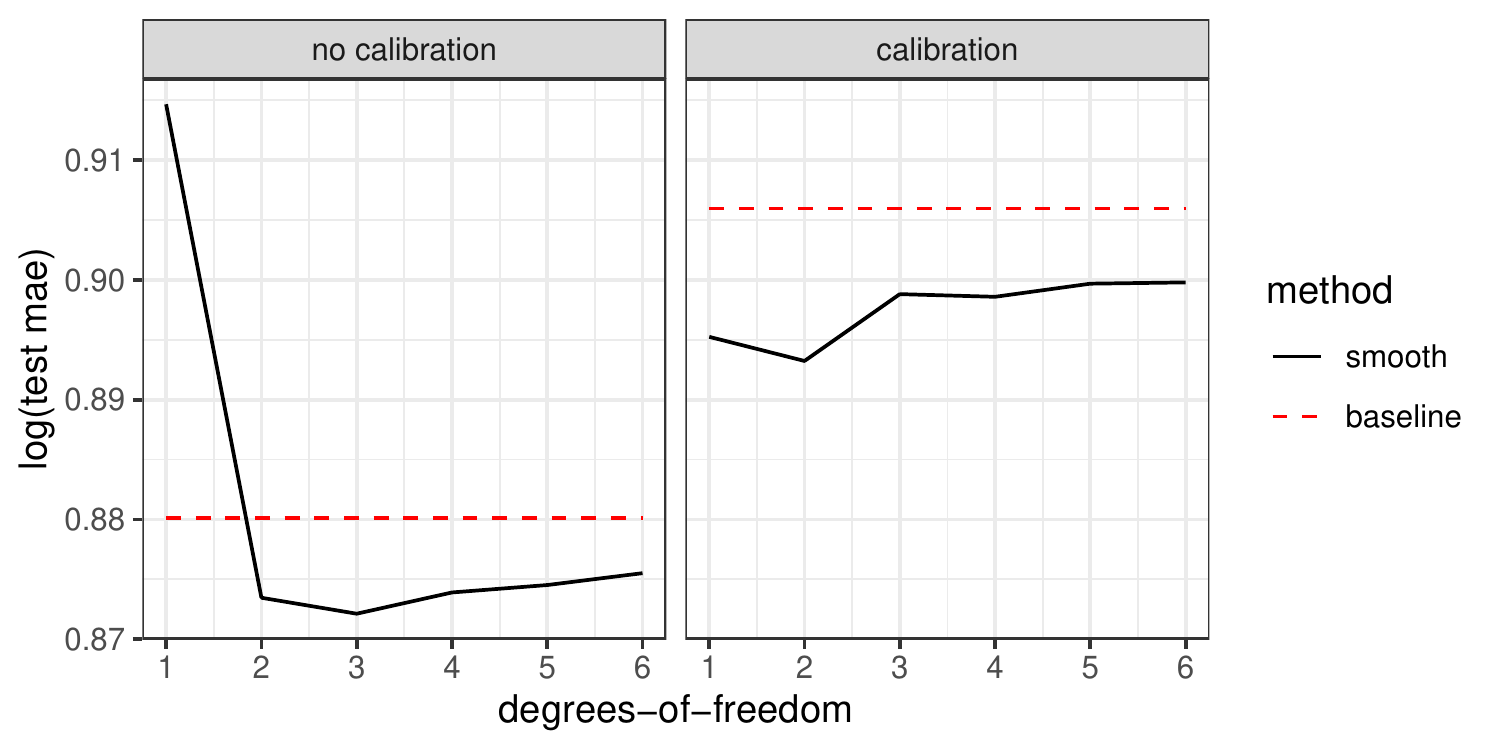}  
\end{subfigure}

\hfill
    
\hfill

\begin{subfigure}{\textwidth}
  \centering
  \includegraphics[width=\linewidth]{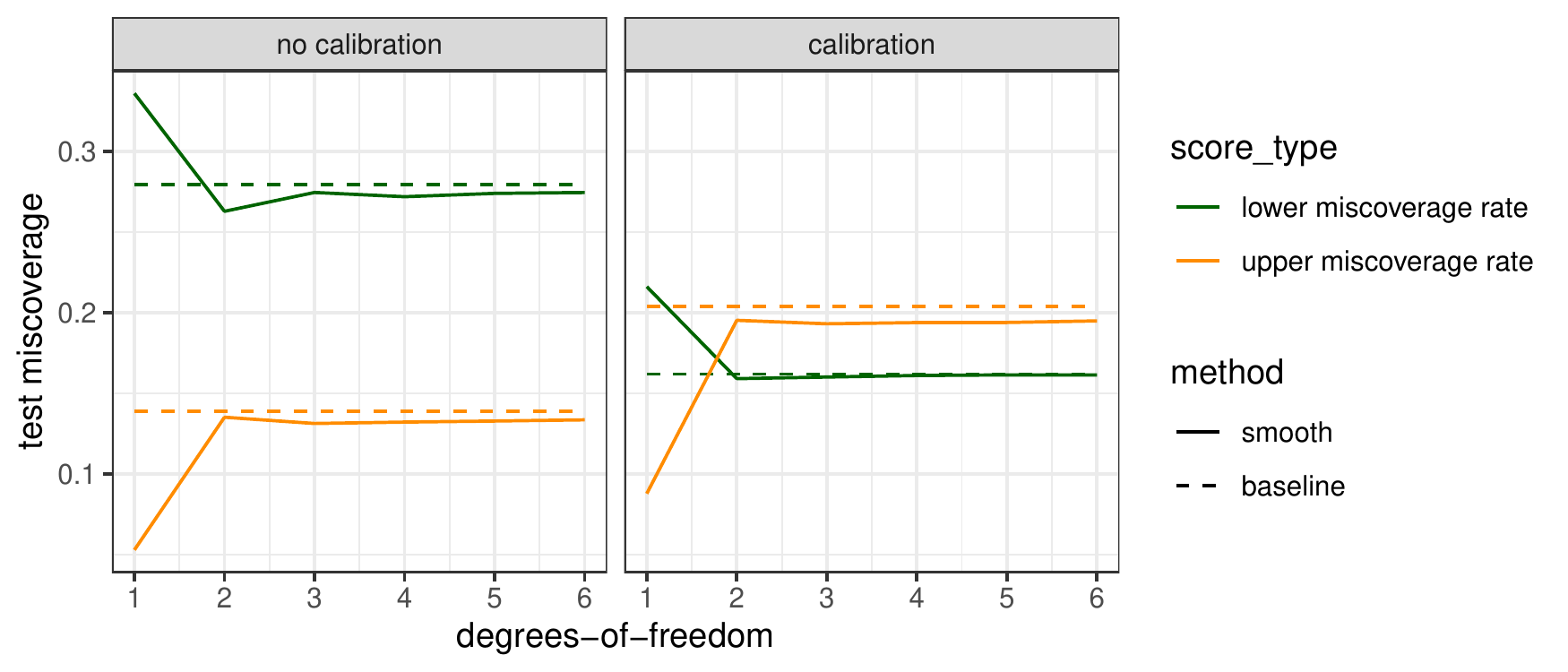}
\end{subfigure}

\hfill

\hfill

\caption{Comparison of the test performance of the baseline and smooth QMPF models for forecasting COVID-19. The plot represents the performance scores produced by the baseline model (dashed line) and the smooth models with different degrees-of-freedom (solid line). The upper plot shows the MAE score whereas the bottom plot shows the upper (orange) and lower (green) miscoverage rates. The target miscoverage rate is $20\%$. The left panel of each plot shows the performance of QMPF before conformal calibration, whereas the right panel represents the calibrated test scores.  The plot demonstrates improved performance of the smooth model relative to the baseline.}
\label{fig:qr:score:vs:df}
\end{figure}

We test both baseline and smooth QMPF techniques on the same COVIDcast data. We restrict our investigation only to the second scenario with partially observed responses.
In our experiments we use three quantiles: $\tau = 0.5$ that corresponds to the predicted median value of cases and $\tau = 0.2, 0.8$ that we use to compute lower and upper bounds for the predicted intervals. For each $\tau$ we solve the QMPF optimization problem and calculate the resulting fit according to (\ref{eq:fit}), which we hereafter denote by $\widehat Y^\tau_i(t+a)$. We denote by $M$ the number of observed responses, i.e. 
$M = \sum_{i = 1}^n \sum_{t\in T} |A_i(t)|,$ and track three performance metrics:
\begin{align*}
    \mbox{ mean absolute error (MAE)} &= \frac 1 {M} \sum_{i = 1}^n \sum_{t\in T} \sum_{a \in A_i(t)} \big|Y_{i}(t+a) - \widehat Y^{0.5}_{i}(t+a)\big|,\\
    \mbox{lower miscoverage rate (LMR)} &= \frac 1{M} \sum_{i = 1}^n \sum_{t\in T} \sum_{a \in A_i(t)} \indicator{Y_{i}(t+a) < \widehat Y^{0.2}_{i}(t+a)},\\
    \mbox{upper miscoverage rate (UMR)}
    &= \frac 1{M} \sum_{i = 1}^n \sum_{t\in T} \sum_{a \in A_i(t)} \indicator{Y_{i}(t+a) > \widehat Y^{0.8}_{i}(t+a)}.
\end{align*}
Here $\indicator{\mathcal{B}}$ refers to the indicator function, taking the value 1 on the event $\mathcal{B}$ and 0 otherwise. We evaluate these three metrics on the test set and present the results in Figure \ref{fig:qr:score:vs:df}. According to the upper left panel, the smooth model with the lowest MAE score has $d=3$ degrees-of-freedom. Despite implying that cases should be forecast in a simplistic quadratic fashion, it outperforms the baseline model in terms of MAE. In the bottom left panel of the plot we show the miscoverage rates obtained by 0.2 (green) and 0.8 (orange) quantiles. From this plot we can conclude that smoothing not only decreases the mean absolute error, but also can be helpful in improving the QMPF coverage, though this improvement is slight.

Analogously to Figure \ref{fig:fits}, we also examine the fitted values obtained by the baseline and the smooth QMPF model with three degrees-of-freedom. For simplicity, in Figure \ref{fig:qr:score:vs:df} we present the forecasted values for one timestamp (i.e. October 2) and the twenty counties with the highest average rate of cases. From the plot we can infer that for some counties, e.g.\ 01003 or 01097, smoothing can improve the prediction accuracy, although for others, e.g.\ 45035 or 45063, the difference is not considerable. 



\begin{figure}[p]
    \centering
    \includegraphics[width = \textwidth]{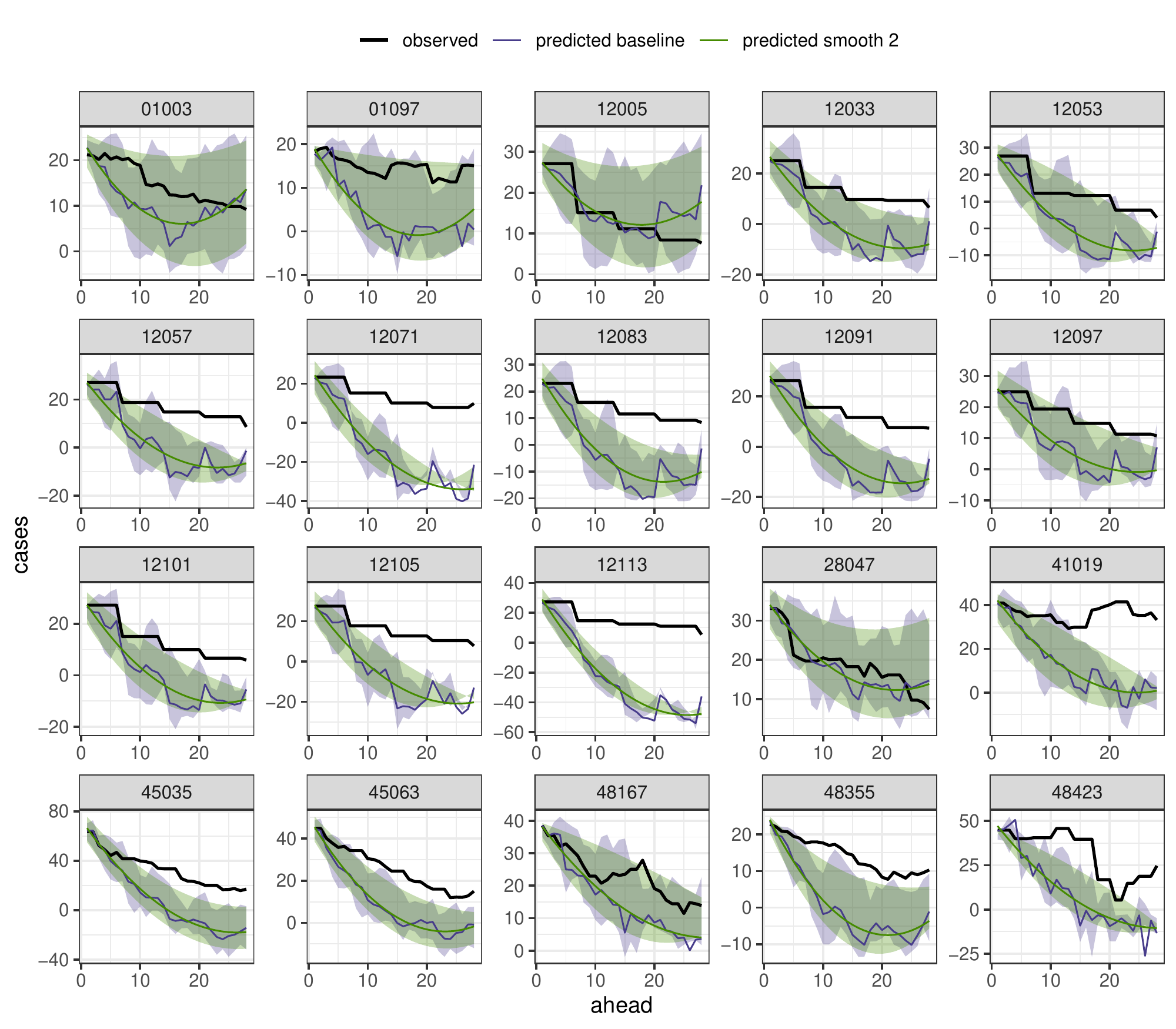}
    \caption{Comparison of the test predictions of the baseline and smooth QMPF models for forecasting COVID-19. The plot displays the out-of-sample fits produced by the baseline (purple) and the best smooth model with $d = 3$ (green). The fits are presented only for October 2. The bold black line shows the true observed newly reported cases, whereas predicted values are represented by thin colored lines. The prediction interval obtained by $0.2$ and $0.8$ quantiles is also displayed (shaded region).}
    \label{fig:qr:fits}
\end{figure}

\section{Conformal calibration}
\label{calibration}

Note that for both $\tau=0.2, 0.8$ quantiles we expect to observe miscoverage of about 20\%. Thus, QMPF models demonstrate mild undercoverage by the lower bound and more sever overcoverage by the upper one (see the left bottom panel of Figure \ref{fig:qr:score:vs:df}). In this section we apply calibration to the QR model which allows us to improve the coverage on the test set.

Conformal quantile regression is a method for constructing prediction intervals that, without making distributional assumptions, helps achieve proper coverage in finite samples (see, for example, \citep{romano2019}).
The idea of this technique is to perform calibration of predicted values on some independent set. Thus, as a first step we split out training data into two parts: we refit the model on the first part and use the second one to calibrate the predicted cases. To reduce the correlation between these parts, we hold out four weeks of the most recent timestamps from $T_{\textrm{train}}$ for calibration, i.e. 
\begin{align*}
T_{\textrm{train}} &= T_{\textrm{train}}^{\textrm{fit}}\bigcup T_{\textrm{train}}^{\textrm{cal}},\\
T_{\textrm{train}}^{\textrm{fit}} &= \{\mbox{10-Jul-2021, 11-Jul-2021, \ldots, 3-Sep-2021}\},\\
T_{\textrm{train}}^{\textrm{cal}} &= \{\mbox{4-Sep-2021, 5-Sep-2021, \ldots, 1-Oct-2021}\}.
\end{align*}
After fitting QMPF models on $T_{\textrm{train}}^{\textrm{fit}}$ we use the resulting coefficients to evaluate the fits $\widehat Y^{\tau}_i(t+a)$ as well as the upper and lower errors
\begin{align*}
 E_i^{0.2}(t+a) = \widehat Y^{0.2}_i(t+a) - Y_i(t+a),\\   
 E_i^{0.8}(t+a) = Y_i(t+a) -  \widehat Y^{0.8}_i(t+a).
\end{align*}
Then, we usey $T_{\textrm{train}}^{\textrm{cal}}$ to calculate the margins  
\begin{align*}
Q^{0.2} &= \mbox{ 0.8-th empirical quantile of } \{E_i^{0.2}(t+a): i\in[n],~a\in A,~t \in T_{\textrm{train}}^{\textrm{cal}}\},\\
Q^{0.8} &= \mbox{ 0.8-th empirical quantile of } \{E_i^{0.8}(t+a):~i\in[n],~a\in A,~t \in T_{\textrm{train}}^{\textrm{cal}}\},
\end{align*}
and replace the original prediction interval $[\widehat Y^{0.2}_i(t+a),\ \widehat Y^{0.8}_i(t+a)]$
with its calibrated version
$[\widehat Y^{0.2}_i(t+a)-Q^{0.2},\ \widehat Y^{0.8}_i(t+a)+Q^{0.8}]$.

We display the performance of QMPF after calibration in the right panel of Figure \ref{fig:qr:score:vs:df}. As one can see from the bottom right panel of the plot, the procedure considerably improves the coverage, which is now much closer to the reference 20\%. According to the upper right panel, the optimal smooth model has $d = 2$ degrees-of-freedom, suggesting forecasting a linear trend for cases. Finally, analyzing both panels, we conclude that, even for calibrated models, the smoothing technique still outperforms the baseline method on the test set. 

\section{Discussion}
In this paper, we proposed a time-series forecasting approach intended to predict multiple ``ahead'' values of the signal simultaneously. The baseline method, commonly used in the literature, suggests treating each ahead value independently, thereby fitting several separate models. On the contrary, the smooth MPF technique takes into account that the same signal measured at different time points in the forecasting model. It assumes that the model coefficients depend smoothly on time, thereby forecasting multiple ahead values with a single smooth curve. We develop the proposed approach in a least-squares framework, which can be handled easily by multiple linear regression. Subsequently, we extend the methodology to forecasting the prediction intervals via quantile regression. We illustrate the benefits of smoothing in the context of multi-period forecasting through a small simulation as well as on an example using county-level COVID-19 incident cases.

There remains additional opportunity for future work.
In the current study, we consider a limited set of predictors: cases, estimated percentage of people experiencing COVID-like illness, and the proportion of people reporting illness in their local community. One interesting direction would be to extend this set and include additional indicators from the COVIDcast  database such as social behavior or mobility data. From the methodological point of view, this would require us to develop an efficient way to combine smooth multi-period forecasting with regularization. For instance, smooth structure in the coefficients can be handled by group-type penalties such as group-lasso.

\section{Software}
\label{sec5}
The code for the proposed methods is available from
\href{https://github.com/ElenaTuzhilina/MPF}{https://github.com/ElenaTuzhilina/MPF}.

\section*{Funding}

Elena Tuzhilina was supported by Stanford Data Science Institute. 
Trevor J. Hastie was partially supported by grants DMS-1407548 and IIS
1837931 from the National Science Foundation, and grant 5R01 EB
001988-21 from the National Institutes of Health.
Robert Tibsirani was supported by the National Institutes of Health (5R01 EB001988-16) and the National Science Foundation (19 DMS1208164).
Daniel J. McDonald was supported by the National Sciences and Engineering Research Council of Canada  (RGPIN- 2021-02618).

\section*{Acknowledgments}
The authors thank the Delphi Research Group, especially, Larry Wasserman, Valérie Ventura, Collin Politsch, Logan Brooks, Jed Grabman and Mike O'Brien for very helpful comments and suggestions.

\noindent
{\it Conflict of Interest}: None declared.

\bibliographystyle{unsrt} 
\bibliography{references}

\end{document}